\documentclass[twocolumn]{aastex631}

\shorttitle{SpinSpotter}
\shortauthors{Holcomb et al.}
\graphicspath{{./}{figures/}}
\newcommand{\tess}{\emph{TESS}\xspace}
\newcommand{\spinspotter}{\texttt{SpinSpotter}\xspace}

\usepackage{graphicx}
\usepackage{natbib}
\usepackage{xcolor}
\usepackage{multirow}
\usepackage{xspace}
\usepackage{amsmath}

\usepackage{txfonts}
\begin{document}

% https://www.overleaf.com/project/5ef1253868892600013c0b11   
\title{SpinSpotter: An Automated Algorithm for Identifying Stellar Rotation Periods With Autocorrelation Analysis}

\author[0000-0002-5034-9476]{Rae J. Holcomb}
\affiliation{Department of Physics \& Astronomy, University of California Irvine, Irvine, CA 92697, USA}

\author[0000-0003-0149-9678]{Paul Robertson}
\affiliation{Department of Physics \& Astronomy, University of California Irvine, Irvine, CA 92697, USA}

\author[0000-0002-5380-549X]{Patrick Hartigan}
\affiliation{Physics and Astronomy Department, Rice University, Houston, TX 77005, USA}

\author[0000-0002-0582-1751]{Ryan J. Oelkers}
\affiliation{Charles R. and Judith G. Munnerlyn Astronomical Laboratory, Texas A\&M University, College Station, TX 77832, USA  }
\affiliation{George P. and Cynthia W. Mitchell Institute for Fundamental Physics and Astronomy, Department of Physics and Astronomy, Texas A\&M University, College Station, TX 77843, USA }

\author{Caleb Robinson}
\affiliation{Physics and Astronomy Department, Rice University, Houston, TX 77005, USA}



\begin{abstract}

\spinspotter is a robust and automated algorithm designed to extract stellar rotation periods from large photometric datasets with minimal supervision. Our approach uses the autocorrelation function (ACF) to identify stellar rotation periods up to one-third the observational baseline of the data. Our algorithm also provides a suite of diagnostics that describe the features in the ACF, which allows the user to fine-tune the tolerance with which to accept a period detection. We apply it to approximately 130,000 main-sequence stars observed by the Transiting Exoplanet Survey Satellite (TESS) at 2-minute cadence during Sectors 1-26, and identify rotation periods for 13,504 stars ranging from 0.4 to 14 days. We demonstrate good agreement between our sample and known values from the literature and note key differences between our population of rotators and those previously identified in the Kepler field, most notably a large population of fast-rotating M dwarfs. Our sample of rotating stars provides a data set with coverage of nearly the entire sky that can be used as a basis for future gyrochronological studies, and, when combined with proper motions and distances from Gaia, to search for regions with high densities of young stars, thus identifying areas of recent star formation and undiscovered moving group members. Our algorithm is publicly available for download and use on GitHub.

\end{abstract}


%
%-------------------------------------------------------------------


\section{Introduction} \label{sec:intro}

Understanding the rotation of stars has long been foundational to characterizing stellar formation, magnetism, and evolution. In the past decade, the study of stellar rotation has been revolutionized by large-scale photometric surveys, particularly the Kepler and K2 missions, and provided the first statistical samples of photometrically measured stellar rotation periods. In 2020, the Transiting Exoplanet Survey Satellite (\tess) completed its primary mission of photometrically surveying bright stars across more than 90\% of the night sky, allowing the first all-sky statistical search for bright rotating stars among hundreds of thousands of targets. To ingest such a wealth of data, we present an automated algorithm that uses autocorrelation analysis to identify the rotation periods of stars with high cadence photometric data. We also present a catalog of stellar rotation periods with $P_{rot} < 14$ days recovered from the sample of 2-minute cadence \tess targets. 

Historically, early rotation measurements were performed through spectroscopic measurements of rotational broadening or through long term ground-based photometric surveys \citep[e.g.][]{mtwilson}. The basis of the photometric approach is to detect the brightness variations from spot complexes on the photosphere. Active regions on a star’s surface typically appear darker than the surrounding surface; thus, as a star rotates this creates quasi-periodic modulations in the star’s brightness as the spots move in and out of the observer’s line of sight. 

The advent of wide field space-based photometric surveys, such as the Kepler mission \citep{borucki2010_keplermissionpaper}, ushered in a new era of rotation studies. Large scale, consistent photometric monitoring of thousands of targets allowed the first statistical-scale rotation surveys based on measuring periodic brightness modulations caused by spots on the surface of rotating stars \citep[e.g.][]{affer2012,harrison2012,aigrain2015}. In particular, the work of \citet{McQuillan2013} demonstrated the power of using the autocorrelation function (ACF) to detect rotation periods, and soon after \citet{McQuillan2014} published the largest sample of measured stellar rotation periods ever, reporting 34,030 rotation periods from a sample of 133,030 Kepler targets. They demonstrated the advantages of the ACF approach over Fourier-based methods for recovering periodic brightness modulations that may vary in shape and amplitude over time as spots form, evolve, and dissipate on the photosphere of a star. Since then, other papers have further refined the ACF method. In 2015, \citet{aigrain2015} performed a comparative analysis of several different methods for recovering stellar rotation from Kepler light curves, including those based on Lomb-Scargle periodograms, autocorrelation, and wavelet decomposition, and concluded that among the methods investigated, ACF-based approaches were most conducive to automated analysis. Thereafter, a variant of the ACF method was developed by \citet{saylor2018} that was focused on recovery specifically of fast rotators in the Kepler sample. 

The \tess satellite launched in 2018 with the primary mission of surveying more than 80\% of the sky across two years to search for transiting exoplanets \citep{ricker2015}. It observed over 200,000 of the brightest stars at a high 2-minute cadence, and millions more at a slower 30-minute cadence. While its primary mission is to discover nearby transiting exoplanets, this survey also provides ideal data to search for fast-rotating stars across the majority of the night sky. Work on this has already begun, such as \citet{canto2020} investigating the rotational properties of \tess stars hosting candidate transiting planets, \citet{ramsay2020} searching specifically for ultra-fast rotating M dwarfs in the southern \tess field, and \citet{claytor2021} applying new deep learning techniques to the \tess sample.

Stellar rotation is of particular interest for studying the evolution of stars during their early lifetimes and evolution onto the main sequence. Solar-type stars have long been thought to converge to a tight age-period relationship once they evolve onto the main sequence \citep{skumanich1972}, giving rise to the field of gyrochronology, which sought to establish well described relationships between stellar age and rotation. Considerable work has been put into showing the ages of young stellar associations can be derived from rotation periods of their members \citep[e.g.][]{Agueros2018,Curtis2020,spada2020}, though age determination of field stars through gyrochronology remains elusive. Additionally, the period-age relationship for young late-type dwarfs remains particularly poorly constrained, as these stars will maintain their rapid rotation for far longer than their Solar-like counterparts and then undergo a rapid transition onto the slow rotator track, though the mechanism driving this remains an open question \citep{newton2016}. According to gyrochronological models, fast rotation is often invoked as a signature of youth. However, analysis of fast rotation in Kepler field stars with Gaia parallaxes has thrown doubt on this assumption, suggesting that many fast rotating field stars may be tidally synchronized binaries \citep{simonian2019}. Expanding our collection of known rapid rotators to encompass the whole sky will allow new investigations into these topics.

In this paper, we present two products. First we describe \texttt{SpinSpotter}, a Python-based algorithm that searches for evidence of stellar rotation in densely-sampled photometric time series, such as those provided by the \tess and Kepler missions. Second, we applied \spinspotter to all main sequence stars observed with a 2-minute cadence from the \tess primary mission (Sectors 1-26), both to demonstrate the effectiveness of our algorithm and to create a catalogue of rotating stars with rotation periods of two weeks or less that may be used as a basis for future studies of gyrochronology, stellar activity, and more. Section \ref{sec:algorithm description} describes the workings of our algorithm, Section \ref{sec:rotators in TESS sample} explains how we applied it to the \tess sample, Section \ref{sec:discussion} discusses the significance of our results, and Section \ref{sec:conclusion} presents our conclusions.

%--------------------------------------------------------------------
\section{Algorithm Description} \label{sec:algorithm description}

This section presents the methodology of the \texttt{SpinSpotter} code package, which is available for download and use at \url{https://github.com/rae-holcomb/SpinSpotter}. The algorithm builds on the techniques developed in \citet{McQuillan2013} and \citet{saylor2018} and aims to provide a tool that can be flexibly applied to photometric survey data in order to identify rotating stars from a large sample in a fast, automated way. As input, it takes photometric light curves (LCs) with densely-sampled data, such as those provided by the \tess or Kepler missions. As output, it provides the stellar rotation period if one is identified, as well as a collection of descriptive parameters which characterize various features of the light curve and its ACF, and can be used to tune the sensitivity of the period selection.

\begin{figure*}
\centering
\includegraphics[width=\textwidth]{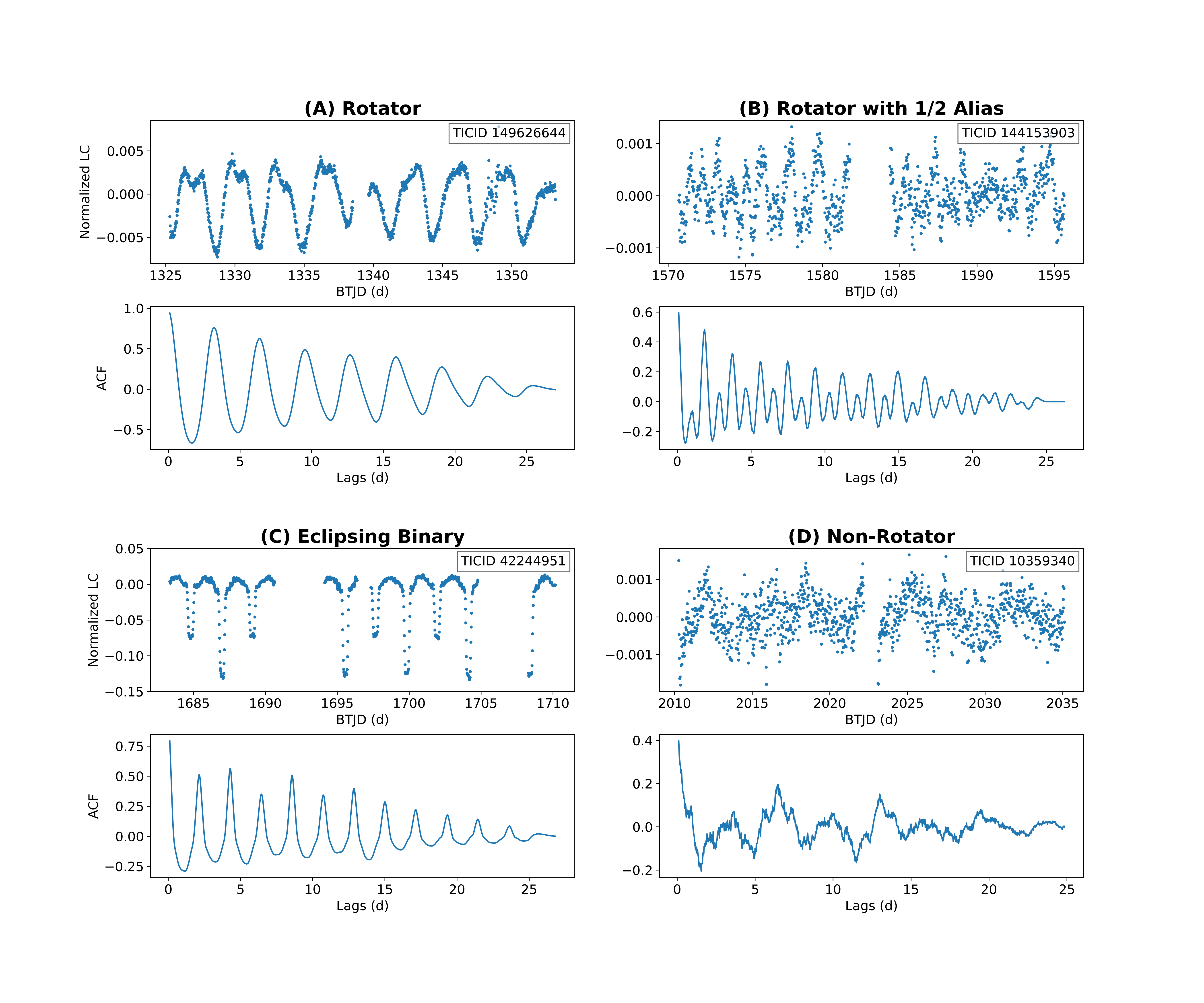}
\caption{Examples of LCs and ACFs for illustrative cases of a rotating star (A), a rotating star with spots in opposite hemispheres thus creating alias peaks at half values of the rotation period (B), an eclipsing binary system (C), and a star with no clear evidence of rotation (D). For the case with spots in opposite hemispheres (B), note the alternating low-high pattern of peaks in the ACF. The even numbered peaks, which are taller, correspond to the true rotation period.}
          \label{fig:lc_example_lots}%
\end{figure*}

\begin{figure*}
\centering
\includegraphics[width=1\textwidth]{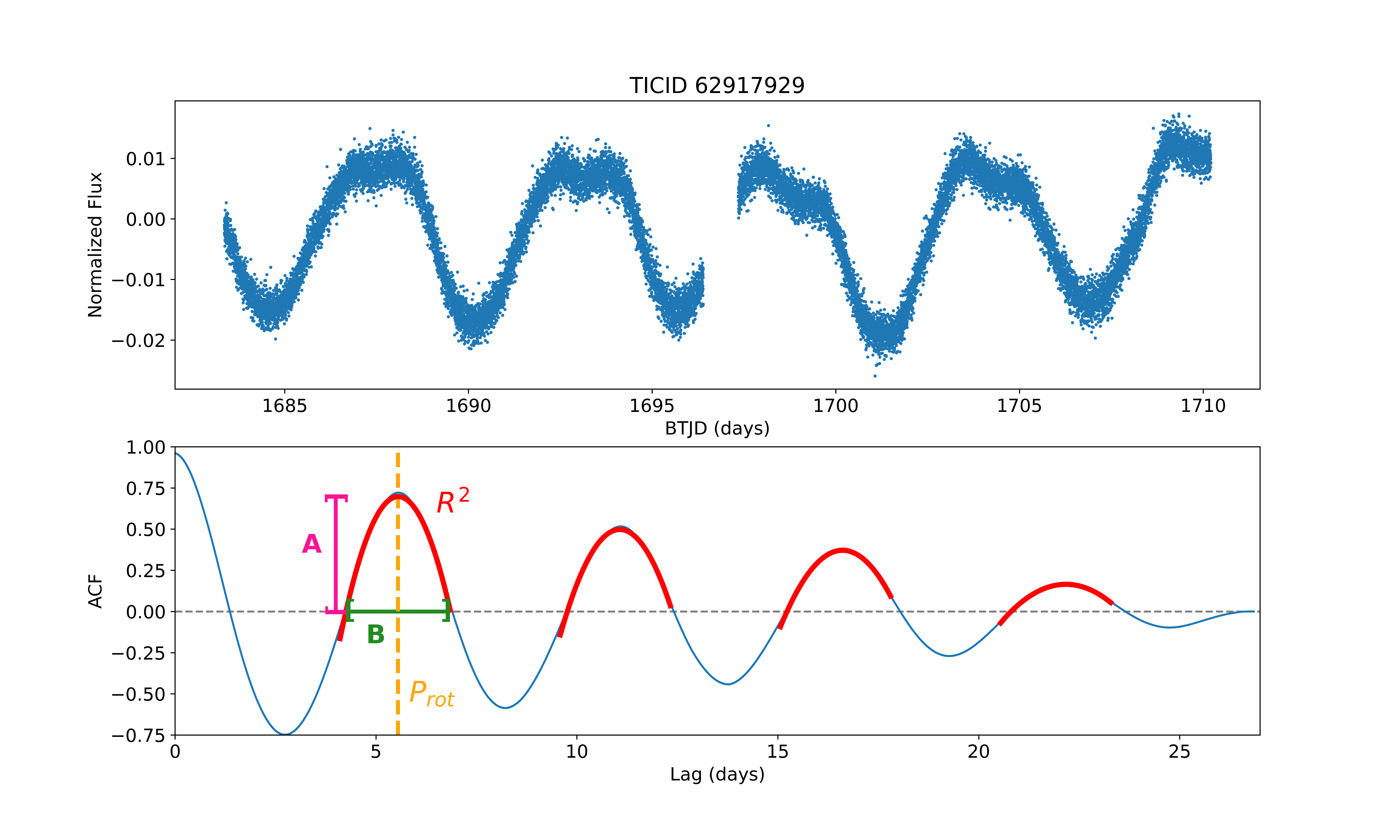}
\caption{An example of the \tess light curve (top) and ACF (bottom) of a star that displays clear evidence of rotation at 5.8 days. The periodicity causes peaks in the ACF to occur at integer multiples of the lag associated with the rotation period. We fitted parabolas around each of these peaks (red) and used the height of the peak vertex (A, pink), the width of the peak at the zero crossing (B, green), and the $R^2$ statistic for the parabolic model (red) as parameters to accept or reject the rotation period associated with the location of the peak vertex ($P_{rot}$, yellow).}
          \label{fig:acf_param_plot}%
\end{figure*}

\texttt{SpinSpotter} performs the following steps. For each star, steps 1-6 are performed on each sector of data individually and also on the combined multisector light curve.

\begin{enumerate}
\item Normalize each sector of data and combine into a single multisector light curve. 
\item Calculate the ACF.
\item Select the highest peak in the fast Fourier transform (FFT) of the ACF to identify an initial estimate for rotation period.
\item Fit parabolas to the peaks in regions of the ACF corresponding to integer multiples of the initial estimate of the rotation period.
\item From the fit parameters, calculate the descriptive parameters $A_k$, $B_k$, $R^2_k$, and $P_k$ for each  $k$th peak in the ACF. At this stage, stars may be flagged as showing a factor of 2 degeneracy in the rotation period identified.
\item Accept or reject the rotation period found for the candidate based on whether the descriptive parameters pass the selection criteria, and how many sectors the rotation period was identified in. 
\end{enumerate}

\emph{1. Normalize and stitch the LC.} We begin by normalizing and preparing the data for analysis. We divide the light curve by its median value and subtract 1 so that the data are centered around zero. If the time series is observed across multiple segments of time, such as \tess sectors or Kepler quarters, this normalization is performed separately on each segment before combination. Next, we stitch the segments together and bin the time series to the desired cadence. After stitching, we subtract the mean of the entire light curve to negate any offsets introduced by the stitching or binning process. Finally, we fill large gaps, such as those introduced by \tess data downlinks and sectors when target is unobserved, with zeros. The process of filling these gaps with zero may affect the ACF for lags longer than the baseline of a single segment of data; however, we only consider lags on timescales shorter than a single segment and thus the effect is negligible.

\emph{2. Calculate the ACF.} We calculate the ACF according to its standard form

\begin{equation}
\label{eqn:acf_eqn}
r_k =  \frac{\sum_{i=1}^{N-k} (x_i - \bar{x}) (x_{i+k} - \bar{x})}{\sum_{i=1}^{N-k} (x_i - \bar{x})^2}
\end{equation}

where $r_k$ is the autocorrelation coefficient at lag $k$, for each point in the time series $x_i$ ($i = 1, …, N$). Each lag $k$ corresponds to $\tau_k = k\Delta t$, where $\Delta t$ is the cadence \citep{shumway2000}.

\emph{3. Take the FFT of the ACF to identify the dominant periodicity in the ACF.} Stellar rotation appears in the ACF as multiple peaks occurring at lags equal to integer multiples of the rotation period, as pictured in Figures \ref{fig:lc_example_lots} and \ref{fig:acf_param_plot}. The number of peaks in the ACF corresponds to the number of cycles of the rotation period captured in a single segment of the data. This means that the ACF of a rotating star shows strong periodic behavior at the same period as the stellar rotation. Thus, to calculate an initial estimate of the rotation period, we take a fast Fourier transform (FFT) of the ACF and use the location of highest FFT peak to guide our period search. This approach works better than simply taking the FFT of the light curve since the shape of the periodic signal tends to be more consistent across multiple cycles than in the original light curve. However, this value by itself cannot be used as an accurate measure of period, since aliases of the rotation signal at late lags in the ACF may drift from integer multiples of the period and thus influence the location of the peak in the FFT, as described in \citet{saylor2018}. Thus, additional steps must be taken to identify the final value of the rotation period.

\emph{4. Fit parabolas to the peaks in the ACF.} We refine the period estimate further by searching for local maxima in the ACF near the value identified from the FFT. We separate the ACF into sections centered around integer multiples of the initial period estimate and fit parabolas to them in order to identify the local maxima in each section, as show in Figure \ref{fig:acf_param_plot}. Parabolas are chosen to provide a simple model for the shape of a smooth peak, whose coefficients can be used to describe the characteristic shape of each peak. We use non-linear least square regression to perform these fits in windows that have width equal to half the initial period estimation, centered on integer multiples of the candidate rotation period. For the $k$th peak, this takes the form

\begin{equation}
y_k = a_k x^2 + b_k x + c_k
\end{equation}

\noindent where $a_k$, $b_k$, and $c_k$ are fitted parameters. 

\emph{5. Calculate the descriptive parameters $P_{rot,k}$, $A_k$, $B_k$, $R_k^2$.} From the fitted parameters of the parabolas, we derive several useful diagnostics that describe the shape and location of the parabola fitted to the peak.

\begin{itemize}
    \item[] \begin{equation}
        P_{rot,k} = - \frac{b_k}{2 a_k}
        \end{equation}
    \item[] \begin{equation}
        A_k = c_k - \frac{b_k^2}{2 a_k}
        \end{equation}
    \item[] \begin{equation}
        B_k = - \frac{ \sqrt{b_k^2 - 4 a_k c_k} }{ a_k}  * \frac{1}{P_{rot}}
        \end{equation}
    \item[] \begin{equation}
        R_k^2 = \textrm{coefficient of determination for the parabola fit}
        \end{equation}
\end{itemize}

\noindent where $P_{rot,k}$ is the location of the parabola peak along the x-axis, $A_k$ is the height of the parabola vertex above the x-axis, and $B_k$ is the width of the parabola at its zero crossing. These expressions correspond to geometric features of the fitted parabola. The coefficient of determination $R_k^2$ is used to assess the goodness of fit of the parabola to the ACF peaks, according to

\begin{equation}
\label{eqn:r2_eqn}
R_k^2 = 1 - \frac{\sum_{i=1} (y_i - \hat{y}_i)^2 }{\sum_{i=1} (y_i - \bar{y})^2}
\end{equation}

\noindent where $y_i$ is the value of the ACF at position $i$, $\hat{y}_i$ is the value of the fitted parabola at position $i$, and $\bar{y}$ is the mean of the ACF in the relevant window. The values of the $P_{rot,k}$, $A_k$, $B_k$, $R_k^2$ are then averaged across each peak in the ACF to produce $P_{rot}$, $A_{avg}$, $B_{avg}$, $R_{avg}^2$. 

In cases where a star is strongly spotted on opposite hemispheres, the light curve will show strong dimming twice per cycle, thus resulting in intermediary peaks in the ACF near the half-period values and creating an alternating high-low pattern in the peaks. Figure \ref{fig:lc_example_lots} shows an example of such a star. To account for these scenarios, we check the height $A_k$ of the first three peaks appearing at the shortest lag times in the ACF. If the second peak is higher than its neighbors by more than 10\%, then we take this as an indication that the star is displaying spots in opposite hemispheres, and thus calculate the rotation period as the average of the vertex location of the even numbered peaks only. The pipeline flags stars where this was applied, and we caution that stars flagged in this way should be checked individually for confirmation that the correct period was selected.

We calculate the period uncertainty with one of two methods. In cases where at least three peaks are identified in the ACF, we take the period uncertainty to be the standard deviation of $P_{rot,k}$, representing the vertex location of each peak in the ACF. When fewer than three peaks are identified, we take the uncertainty to be the average of the half-width at half maximum of the the peaks in the ACF. It should be noted that the latter approach systematically returns larger uncertainties; however, this is appropriate since the rotation period of a star cannot be well known if multiple cycles of the rotation period are not captured in the the light curve.

\emph{6. Apply selection criteria to accept or reject the rotation period.} At this stage, the pipeline has identified a rotation period, period uncertainty, and three additional parameters ($A_{avg}$, $B_{avg}$, and $R_{avg}^2$, describing the peak height, width, and goodness of fit respectively) that describe the ACF peak shape for each star. From these, the user may determine cut off criteria to decide on whether to accept or reject the rotation period found for each star. The best thresholds to use for this may vary depending on the science needs and the input data. In \ref{ssec:Period detection of TESS Rotators}, we describe how we applied this algorithm to the \tess 2-minute cadence data and selected thresholds for our acceptance criteria to identify an all-sky  sample of rotating stars with periods of two weeks or less.

\section{Identifying Rotators in the \tess Sample} \label{sec:rotators in TESS sample}

\subsection{Data Products} \label{ssec:data products}

We used data from Sectors 1-26 of the \tess mission, which is publicly available from the the Mikulski Archive for Space Telescopes (MAST). Each sector has an observational baseline of $\sim27$ days, during which stars are observed at a 2-minute cadence, with a 16 hour gap halfway through each sector due to data downlinks. We used the the PDCSAP\_FLUX light curves provided by the standard \tess pipeline, which corrects for most instrumental and systematic effects \citep{spocpipeline}.

\subsection{Sample Selection} \label{ssec:Sample Selection}

In order to show photometric modulation from photospheric spot complexes, a star must have a convective outer envelope; that is, it must be on the main sequence. To ensure that our targets satisfied this requirement, we applied $T_{\mathit{eff}} - \log g$ cuts recommended by \citet{ciardi2011}, where a star is considered to be a dwarf if its $T_{\mathit{eff}}$ and $\log g$ fall within:

\begin{equation*}
  \log (\textit{g}) \ge
  \left\{
    \begin{aligned}
      & 3.5 & \textrm{if } T_{\mathit{eff}} \ge 6000  \\
      & 4.0 & \textrm{if } T_{\mathit{eff}} \le 4250 \\
      & 5.2 - (2.8 \times 10^{-4}T_\mathit{eff}) & \textrm{if } 4250 < T_{\mathit{eff}} < 6000
    \end{aligned}
  \right.
\end{equation*}

We used $T_{\mathit{eff}}$ and $\log g$ as provided by the \tess Input Catalog v7 \citep{stassun2018}. Any targets with no measured $T_{\mathit{eff}}$ or $\log g$ values were also excluded. This removed 79,087 stars from our initial sample of 215,955 stars. \tess Objects of Interest (TOIs), which are stars that have had planetary transit events identified in their light curves, were included in this analysis, with extra steps taken to mask out potential transit events. Repeated transit events can create narrow peaks in the ACF occurring at integer multiples of the transit period, making it more difficult to isolate the rotation period. Thus, we removed possible transit events by replacing points that occured during a transit (as reported in the TOI list maintained in \citet{TOIs_guerrero2021}) with 0 after the light curve is been normalized, as described in Section \ref{sec:algorithm description}. Since our light curves were normalized to have a mean value of 0, this means that points occurring during these putative transit events did not contribute to the $r_k$ coefficient when calculating the ACF according to Equation \ref{eqn:acf_eqn}.  The TOI list used in this analysis was up to date as of February 4, 2022. 

Eclipsing binary systems and pulsating variables can also create periodic signals not associated with stellar rotation, thereby creating the potential for false positive detections in our pipeline. At the time of this writing, there were no published catalogs of eclipsing binaries (EBs) or pulsating variables (PVs) for the entirety of the \tess field, so additional steps to remove these objects were applied after the primary analysis based on features of the light curve itself. We defined a parameter, $X_{var}$, which is calculated by taking the midpoint between the value at the 95th percentile point and the value at the 5th percentile point of the normalized light curve. The value of $X_{var}$ gives a measure of how well the most extreme variability in the light curve is centered around the mean of the light curve. Since we first normalized the lightcurves to center around 0, a value of $X_{var}$ much larger or smaller than 0 indicated that the most extreme points are highly asymmetric about the mean, which is a feature commonly exhibited in the photometry of eclipsing binaries. We removed any stars with $|X_{var}| > 0.01$ from our analysis, due to their potential binary nature. Most PVs were excluded from the sample by the $T_{\mathit{eff}} - \log g$ cuts that removed evolved stars. To help identify any remaining PVs, we defined a parameter $Y_{var}$ to be the distance between the 5th and 95th percentile points of the normalized light curve, to serve as a measurement of the total variability of the time series. We did not exclude any stars based on the $Y_{var}$ value, though targets with particularly high $Y_{var}$ values, especially those with hotter $T_{\mathit{eff}}$, may warrant additional inspection for evidence of whether their periodic variability arises from rotation or pulsation. After all cuts were applied, the resulting sample contained 136,868 stars with observational baselines ranging from a single sector to 13 sectors if the star was in the continuous viewing zone.

\subsection{Period Detection of TESS Rotators} \label{ssec:Period detection of TESS Rotators}

After applying all cuts, we ran our targets through the \texttt{SpinSpotter} algorithm as described in Section \ref{sec:algorithm description}. We chose to bin the light curves from the 2-minute cadence provided by \tess to a 30 minute cadence, which offered two advantages. First, it reduced the contribution of variability on timescales of less than 30 minutes, resulting in smoother and more easily identifiable peaks in the ACF on timescales of hours to days, which is the region where we most expect to find stellar rotation. Second, it matched our data to the cadence of both the \tess full frame images and the Kepler mission, thus allowing for easier comparisons between our results and other studies. It should be noted that this choice of binning imposed a limit on our sensitivity to ultra-fast rotators, and we did not recover rotation periods of less than 0.4 days in this study, though we have found it to be possible when using shorter binning timescales. \tess full-frame image (FFI) targets were not included in this analysis, as at this time they have been not run through the \tess data reduction pipeline, though they may be in the future. Once this is done, these targets would make an excellent sample to analyze with the \texttt{SpinSpotter} algorithm in future works.

\subsection{Development of Training Data \& Selection Criteria} \label{ssec:Training Data}

\texttt{SpinSpotter} provides descriptive statistics about the peak height, width, and goodness of fit in the ACF, but what values of these statistics indicate credible identification of the rotation period? In order to define acceptance criteria for our sample, we visually inspected 3285 LCs and their corresponding ACFs selected randomly from our sample and assigned them one of the following categorizations:
    
\begin{itemize}
  \item \textbf{Periodic (379 stars)} - periodicity clearly visible in the LC, with matching peaks in the ACF.
  \item \textbf{Possibly Periodic (222 stars)} - some evidence of periodicity in the LC and ACF, but less confident than in the Periodic category.
  \item \textbf{Eclipsing Binary (32 stars)} - LC indicative of being an eclipsing binary system.
  \item \textbf{Non-Periodic (2652 stars)} - no clear evidence of periodicity.
\end{itemize}

We randomly assigned 70\% of these tagged stars to be used for tuning the values of our selection criteria, and the remaining 30\% to be used to verify how well they perform, resulting in a tuning set and a verification set of 2300 and 985 stars respectively. Using the former set, we chose the following selection criteria to accept a period detection as valid:

\begin{itemize}
  \item[] \begin{center} $A_{avg}/B_{avg} > 0.25$ \end{center}
  \item[] \begin{center} $.4 < B_{avg} < 0.6$ \end{center}
  \item[] \begin{center} $R_{avg}^2 > 0.9$ \end{center}
\end{itemize}

For stars with multiple sectors of data, we also required that a valid period be detected in a minimum number of sectors, as well as in the combined multisector light curve. We defined this minimum to be half the total number of sectors for which the star was observed, rounded up. \tess sectors are short and often non-consecutive, such that spot patterns that appeared in one sector of data may no longer be present in later sectors, and the noise level between different sectors of the same target can also vary dramatically \citep{dalba2020}. 

\begin{figure*}
\centering
\includegraphics[width=\textwidth]{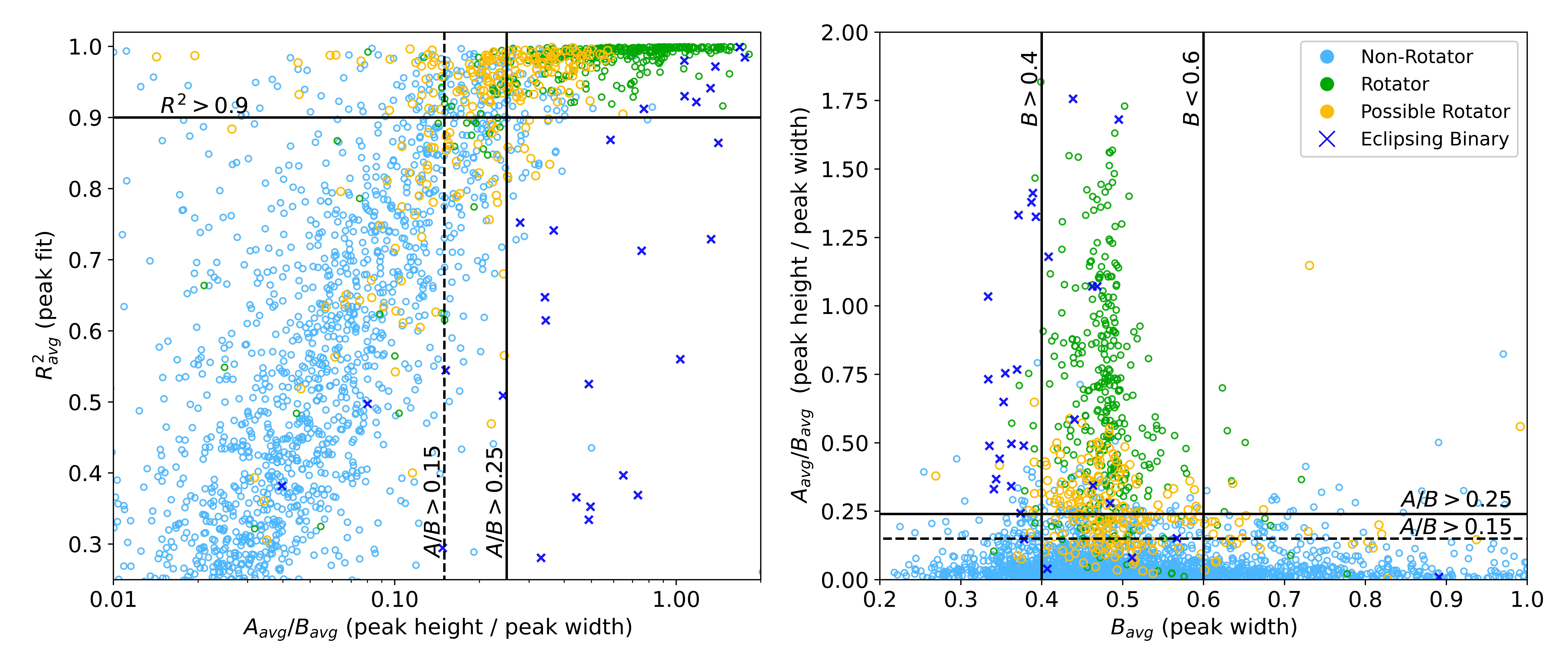}
\caption{Stars from our testing and verification sets, plotted to show our selection criteria. $A_{avg}$ is the ACF peak height, $B_{avg}$ is the fractional peak width, and $R^2_{avg}$ is the goodness of fit parameter of the parabolas. Solid black lines show the values of our selection criteria, dashed lines show suggested values of relaxed criteria that would provide a higher recovery rate of rotators at the expense of a higher false positive rate.}
\label{fig:parameter_criteria_plot}
\end{figure*}

In the tuning set of 2300 stars, these criteria identified 203 stars as rotators, with a false positive rate of 4.93\%. We recovered 62.5\% of stars tagged as rotators and 18.5\% of those marked as possible rotators. We found that the majority of our false positives were identified as having periods longer than 9 days, which matched one third the baseline of a single \tess sector. When the sample was restricted to only accept stars with $P_{rot} < 9$ days, our false positive rate fell to 3.3\% and we recovered 70.1\% of tagged rotators and 29.8\% of tagged possible rotators. Thus, we concluded that our algorithm has limited ability to recover rotation periods that exceed a third of the baseline of one \tess sector, the reasons for which we discuss further in Section \ref{sec:discussion}.

Applying these same conditions to the verification set of 710 stars, we found that we identified 84 stars as rotators, with a false positive rate of 6.17\%. We recovered 53.1\% of stars tagged as rotators and 21.1\% of those marked as possible rotators. Thus, we concluded that the criteria derived from our tuning set were generalizable to the entirety of our sample of \tess rotators. More potential rotators could be recovered by relaxing these requirements. For example, lowering the criteria to $A_{avg}/B_{avg} > 0.15$, resulted in the identification of 236 rotating stars in the tuning set and 93 rotating stars in the verification set, thus recovering 71.3\% of tagged rotators and 36.4\% of possible rotators, though the false positive rate increased to 11.8\%. Visually, this can be seen in Figure \ref{fig:parameter_criteria_plot}, which shows the distribution of our tagged stars across the descriptive parameters $A_{avg}/B_{avg}$, $B_{avg}$, and $R_{avg}^2$. Our selection criteria, visible as black lines on the plot, effectively separate rotators from non-rotators.

\section{Discussion} \label{sec:discussion}

\begin{figure*}
\centering
\includegraphics[width=\textwidth]{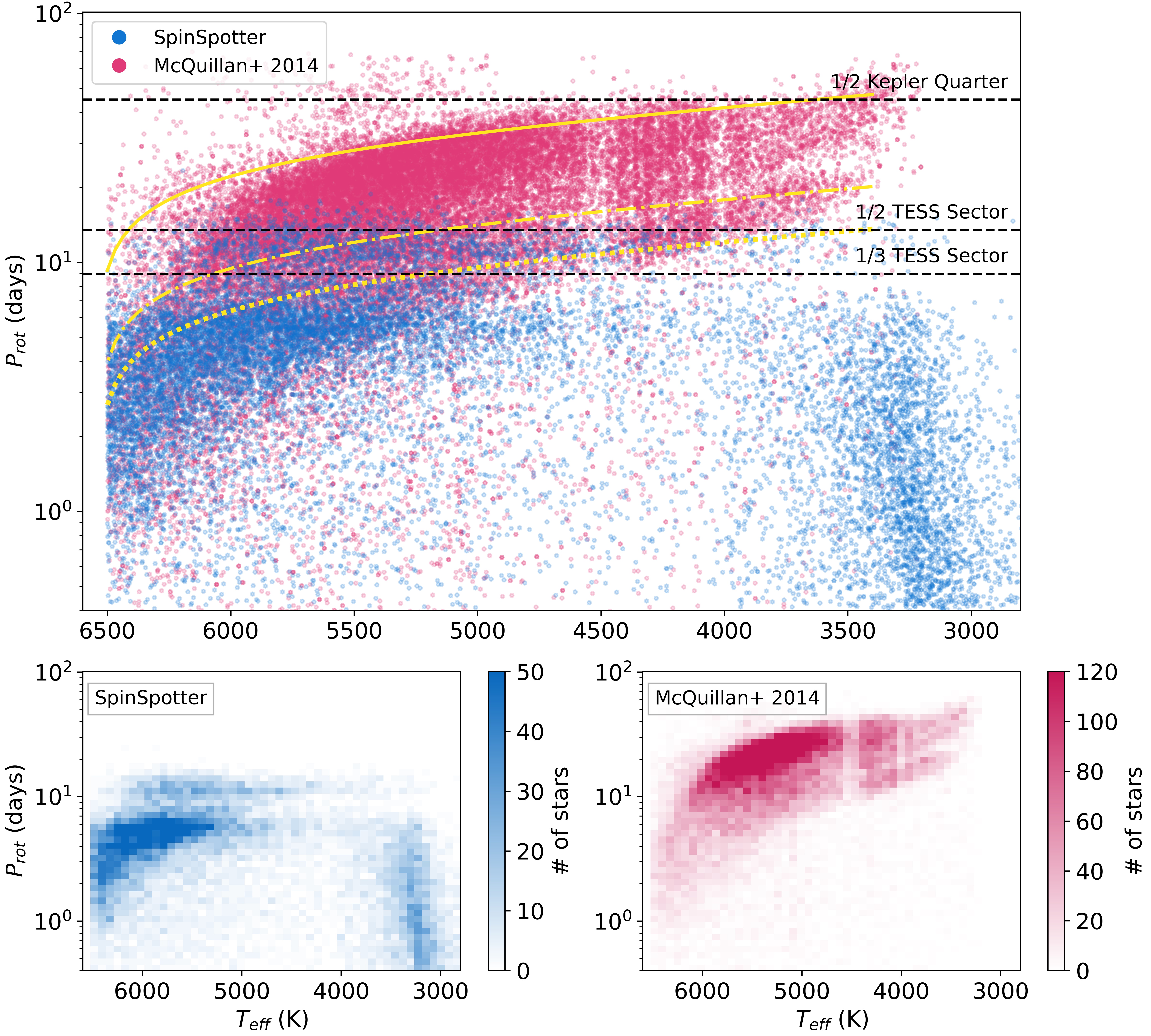}
\caption{\tess rotators identified by our algorithm (blue) alongside those found by \citep{McQuillan2014} (pink) from the Kepler sample. Top: A scatter plot, where each point represents an individual rotating star. Horizontal lines at 9 days, 13.5 days, and 45 days, represent one third and one half of the baseline of a \tess sector, and half the baseline of a Kepler quarter, respectively. Yellow lines show age isochrones from \citep{mamajek2008} at 500 Myr (dotted), 1 Gyr (dash-dotted), and 4 Gyr (solid). Note the large drop off in survey sensitivity for stars with periods longer than 9 days; this reflects our limited sensitivity to longer period rotators due to our short baseline of observations. The increase in rotators with $10 < P_{rot} < 14$ days is dominated by stars flagged as showing half period aliases of the rotation periods. Bottom: Histograms showing the distribution of stars in the \spinspotter (bottom left, blue) and \citet{McQuillan2014} (bottom right, pink) respectively, intended to more clearly show the features of each distribution. Period bins are equal in log space.}
\label{fig:teff_prot_distr}
\end{figure*}

When applied to the whole sample, we found rotation periods for 13,504 stars out of the 136,868 stars that passed our initial cuts in the \tess 2-minute cadence sample. Figure \ref{fig:teff_prot_distr} shows the period and temperature distribution of our sample of rotators compared to those found by \citet{McQuillan2014}. While the two samples show broad agreement, particularly for earlier-type MS stars, there are several salient features of this plot that merit further examination. These are the absence of rotators recovered at longer periods, the anomalously large number of stars identified as rotators in our sample with periods of 10-14 days, and the high number of fast rotating M dwarfs identified in the \tess sample but absent in previous rotation studies of Kepler stars. 

First, we found few stars with rotation periods longer than 9 days. Only 11\% of stars found to have rotation have periods $> 9$ days, which corresponds to approximately one third of the 27-day baseline of a \tess sector. Two effects make these longer period targets difficult to detect. First, in order to identify a rotation period from the ACF, there must be multiple cycles of the rotation period visible, meaning that the star must have been observed for multiple cycles of its rotation period. Secondly, the PDCSAP\_FLUX from \tess is heavily detrended on long time scales, which effectively removes signals on time scales comparable to a sector or longer. This means that even for stars with multiple sectors of data, evidence of rotation may have been removed by the post-processing pipeline. We surmise that it would be possible to recover rotation periods for some of these stars with our algorithm from the \tess sample by building a custom pipeline that extracted light curves from \tess FFIs that preserved such long period signals, but this is beyond the scope of this paper. Additionally, the number of false positive detections among our training data increases drastically above this cut off. Because of this, we caution investigators to treat rotation periods that exceed one third of the baseline of the data with skepticism and recommend additional checks either by eye or with other methods before using them. 

The second feature of interest is the apparent prevalence of stars with $P_{rot}$ between 11 and 14 days. This is not reflective of the underlying stellar population. Our algorithm flags stars where the even numbered peaks in the ACF exceed the height of the odd numbered peaks, since this is often indicative of a star with spots in opposite hemispheres, leading to additional peaks at 1/2 aliases of the rotation period. Stars with this configuration have two peaks in the ACF per rotation period, thus creating more repeated features compared to the typical case which has one peak per rotation period in the ACF, thus making it easier for our algorithm to identify long rotators that have this half period alias. Stars with this configuration make up 74\% of rotators that we identified with periods $> 9$ days. However, many are single sector objects, which makes it difficult to discriminate between the half period and full period case, since other than ultra-fast rotators, most stars will only detect a handful of cycles of the rotation. We flagged stars where the discrepancy in peak height between odd and even peaks is less than 10\% and recommend that these cases be treated on a case by case basis. It is likely that some of the 11-14 day stars have been incorrectly identified as the half-period case and are in fact stars with periods of 5-7 days, which may contribute to the overdensity between 11-14 days in Figure \ref{fig:teff_prot_distr}.

We also see a slight overdensity of stars with rotation periods near 5 days. We attribute this to two main effects. For stars near and above Solar temperatures, we take this overdensity to be representative of the underlying population, as it tracks well with the increased number of rotators found in that region in \citet{McQuillan2014} and other Kepler rotation studies. However, the overdensity persists at cooler temperatures, diverging from the \citet{McQuillan2014} distribution. Based on individual inspection of a random selection of these stars, we suspect that this may be due to uncorrected systematics associated with the PDCSAP\_FLUX. Future works will explore this further, both by applying \spinspotter to datasets other than the PDCSAP\_FLUX to see if this feature persists when a different lightcurve reduction pipeline is used, and comparison to the period distribution of \tess stars as measured by other period-finding methods. 

Finally, our algorithm identifies a large population of fast rotating M dwarfs that is not present in Kepler and K2 rotation surveys such as \citet{McQuillan2014} and \citet{Reinhold2020}. We attribute this to differences in the survey structure of the two missions. The \tess~2-minute cadence sample prioritizes bright, nearby stars, thus leading to a greater inclusion of late M dwarfs in the sample. In contrast, the Kepler mission prioritized Sunlike stars, leading to an exclusion of many M dwarfs from its target list. This phenomenon has also been documented in other papers that investigate the \tess population, such as \citet{ramsay2020}.

The $\sim$9 day limit on our detection sensitivity with \tess 2-minute cadence data has interesting implications for the science cases to which this sample is relevant, particularly when contrasted with other large-scale rotation studies. Figure \ref{fig:teff_prot_distr} shows the temperature-period distribution of our sample of \tess rotators overlaid with those identified in the Kepler sample by \citep{McQuillan2014}. For Sunlike stars, our population looks similar to the \citet{McQuillan2014} sample, but two key features stand out. First, note the appearance of our detection limit at $\sim$9 days, above which we identify very few rotators. Because of this, our sample of rotators will be most useful in studying very young stars that have not yet spun down to lower rotation speeds. This makes our sample of high interest to those studying young stellar objects and open clusters.

\subsection{Comparison to Previous Rotation Studies} \label{previous_studies}

We compared the performance of our algorithm to previous surveys of stellar rotation. Numerous large scale rotation period studies have been performed on the Kepler sample; however, the Kepler field covers only a small portion of the sky compared to \tess, and targeted many stars dimmer than those prioritized in the \tess 2-minute cadence sample, leading to a limited amount of crossover between the two populations. To expand coverage of the sky, we also compared our results to studies of young stellar objects (YSOs) and open cluster members. 
    
Many of these studies had observational baselines significantly longer than that of a \tess quarter, and thus were sensitive to longer-period rotators. We restricted our comparison only to stars with literature values of $P_{rot} < 14$ days, or half of a single \tess sector, which represents the theoretical limit of the period sensitivity of our data.
    
The methods of each paper are summarized below.

\subsubsection{\citet{McQuillan2014}} \label{mcq2014}
\citet{McQuillan2014} examined $\sim$150,000 main sequence stars in Kepler Q3-Q14, excluding known EBs, Kepler Objects of Interest (KOI), and stars with less than 8 quarters of data. They identified periods by calculating the ACF, identifying local maxima, and taking the period to be the slope of these local maxima as a function of peak number. They reported periods for 34,030 stars ranging from 0.5-60 days. Of these, 29 appear in our sample and have $P_{rot} < 14$ days. This low overlap stems from the overall dimness of stars in the Kepler field compared to those targeted by high cadence \tess observations, as well as the much longer baseline of Kepler quarters, which allowed \citep{McQuillan2014} to identify stars with much longer rotation periods.

\subsubsection{\citet{saylor2018}} \label{saylor2018}
\citet{saylor2018} used K2 light curves to search for fast-rotating ($P_{rot} < 4$ days) GKM dwarfs from the SUPERBLINK high proper motion survey \citep{lepine2018_superblink}. Their method was the most similar to the one we elected to use in the analysis of this paper. \citet{saylor2018} calculated the ACF of the light curve, then took the FFT of the ACF to find an initial estimate of the period, which they then refined by fitting parabolas to the peaks appearing at aliases of that initial estimate. They report 481 rotation periods, of which 19 appear in our sample  and have $P_{rot} < 14$ days.

\subsubsection{\citet{Reinhold2020}} \label{reinhold2020}
\citet{Reinhold2020} examined K2 campaigns 0-18 with a variety of analysis techniques. They applied the LS periodogram, wavelet power spectrum, and ACF, then required good agreement between the periods found by each method. They restricted their period searches to $P_{rot} > 1$ day, and typically resolved periods of 1-44 days. They reported periods for 32,387 stars, of which 88 appear in our sample and have $P_{rot} < 14$ days. The low number of stars appearing in both samples can be attributed to similar reasons as for \citet{McQuillan2014}.

\subsubsection{\citet{messina2010} \& \citet{messina2011}} \label{messina}
\citet{messina2010} investigated young stellar associations within 100 pc, including TW Hydrae, $\beta $ Pictoris, Tucana/Horologium, Columba, Carina, and AB Doradus, supplemented with rotation periods for alpha Persei and the Pleiades. They applied the LS periodogram to time series from the All-Sky Automated Survey \citep[ASAS;][]{pojmanski1997_ASAS,pojmanski2002_ASAS} and reported periods of $<$10 days for 144 stars, which were a combination of newly found rotation periods, updates to known periods, and rotation periods from literature. Of these, 76 appear in our sample. \citet{messina2011} extended the analysis of \citet{messina2010} to include objects in Octans, Argus, $\eta$ Chamaeleontis, and IC 2391 clusters. Similarly to the preceding paper, they applied LS periodograms to photometric time series from ASAS and the Wide Angle Search for Planets \citep[SuperWASP;][]{Pollacco2006_superwasp,butters2010_superwasp} archives and reported rotation periods for 132 stars, which included both newly found rotation periods and values from the literature. 49 of these stars appear in our sample.

\subsubsection{\citet{canto2020}} \label{canto2020}
\citet{canto2020} searched for rotation periods among the 1000 \tess Objects of Interest (TOI) from Sectors 1-22 using a combination of fast Fourier Transform, LS periodograms, wavelet techniques, and visual inspection. They found rotation periods for 163 stars ranging from 1 to 13.5 days, of which they designated 131 as “unambigous” periods and 32 as “dubious” periods. 96 of these stars appear in our sample.

\begin{table*}[t]
% \centering
% \caption{Literature Comparison}
\label{tbl:totalparams}
\resizebox{\textwidth}{!}{
\begin{tabular}{lcccccc}
\hline
% \hline

% \multicolumn{7}{c}{\textbf{Planet Parameters}}     \\
% \hline
\textbf{Paper} & \textbf{Sample} & \textbf{Method} & \textbf{\# of xmatched Targets} & \textbf{\# with $P_{rot}$ Agreement within 10\%} & \textbf{\# with $P_{rot}$ Agreement outside 10\%} & \textbf{\# with No Rotation Found}\\
\hline
\citet{McQuillan2014} & Kepler Main Sequence (MS) & ACF & 29 & 14 & 4 & 11 \\
\citet{saylor2018} & K2 MS & Autocorrelation Function (ACF) & 19 & 12 & 2 & 5 \\
\citet{Reinhold2020} & K2 MS & Lomb-Scargle (LS), ACF, Wavelet & 75 & 26 & 10 & 39  \\
\citet{messina2010} & YSOs within 100 pc & LS & 76 & 44 & 15 & 17 \\
\citet{messina2011} & YSOs beyond 100 pc & LS & 49  & 30 & 7 & 9 \\
\citet{canto2020} & \tess Objects of Interest & Fourier Transform, LS, Wavelet & 96 & 46 & 8 & 42 \\

\hline

% \footnotesize $^*$Equilibrium Temperatures assume zero bond albedo
\end{tabular}
}
\caption{A summary of targets that appear both in our sample and in previous studies of stellar rotation. We list the literature sources, the population of the original studies, the primary method of identifying the period used in the original study, the number of stars that appear in both our analysis and the original study with $P_{rot} < 14$ days, and finally the number of those stars that agreed with our found rotation period to within 10\%, disagreed by more than 10\%, and for which we did not find a rotation period.}
\label {table:lit_table}
\end{table*} \label{table:lit_table}

\subsubsection{Results of Comparison to Previous Studies} \label{previous_studies_results}
\begin{figure*}
\centering
\includegraphics[width=\textwidth]{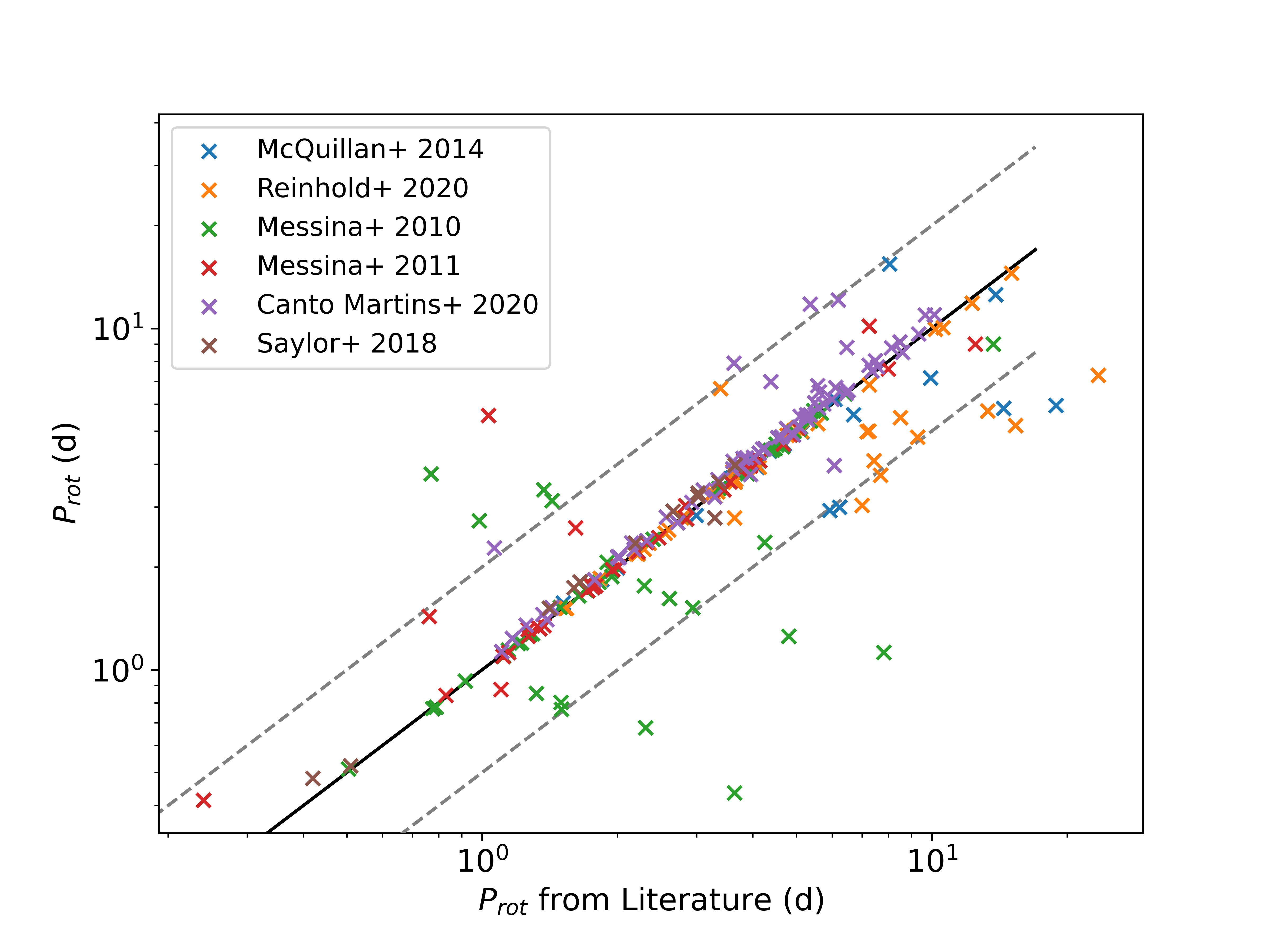}
\caption{A comparison of rotation values as derived by our algorithm (y-axis) and literature values (x-axis). Stars with in agreement with the literature appear on the line y=x (solid black lines). Stars where the period value disagrees between our algorithm and the literature appear near the lines $y=2x$ and $y=x/2$ (grey dashed line). Note the increase in scatter about $y=x$ for stars with $P_{rot} > 9$ days, this reflects our lack of sensitivity to stars periods greater than one third of the data baseline, since multiple cycles of the rotation period must be present in order to form the characteristic peaks in the ACF that are indicative of stellar rotation.}
          \label{fig:lit_comp}%
\end{figure*}

The results of our comparison are given in Table \ref{table:lit_table} and Figure \ref{fig:lit_comp}. Overall, we found good agreement between our rotation periods and those that appear in the literature. Over 75\% of rotation periods identified by our algorithm fell within 10\% of those reported in the literature. Of those that did not agree, we found that a significant portion instead are assigned a value either half or twice that of the literature value. This is visible in Figure \ref{fig:lit_comp}, as clusters of points around the lines $y=2x$ and $y=x/2$. This suggests that both our and previous analyses are identifying signals from the same physical phenomena, but there remains some ambiguity between the true rotation period and the half period aliases that must be addressed individually. Additionally, disagreement is more common among stars with $P_{rot} > 8$ days, once again demonstrating the difficulty of extracting periods that exceed one third of the baseline of the data.

Among the targets of \citet{messina2010}, there are five stars (TIC 383615666, TIC 385012516, TIC 401820823, TIC 425976940, TIC 363782810) with anomalous disagreements between our rotation periods and those reported in the previous study. Upon visual inspection, we find that the period identified by our analysis is consistent with the features of the ACF generated from the \tess time series used in this paper, and the value reported previously is not. \citet{messina2010} reported the periods for four of these five stars with caveats regarding their disagreement with other analyses, thus we conclude with a preference for the rotation periods reported by our analysis.

Our rate of recovering rotators varies greatly from survey to survey, with our highest yields of $>75\%$ among the YSO surveys by \citet{messina2010} and \citep{messina2011}, and the lowest for \citet{canto2020} (56\%) and \citet{Reinhold2020} (48\%). We attribute the “missed” stars to two main effects. First, all these surveys except for Canto 2020 are performed on data that are not concurrent with the \tess mission, meaning that the spots and activity generating the quasi-periodic signals in the light curve may not be present in both periods of observation. Additionally, many of our “missed” stars are observed only for one or two sectors, and this short baseline makes it even more likely that we may have observed the stars during low-activity periods. This is consistent with visual inspection, which reveals that many of the “missed” stars show weak, if any, evidence of their rotation period in their \tess light curve and ACF. 

This explanation does not apply to \citet{canto2020}, who investigated TOIs in \tess sectors 1-22. However, TOIs often show large and frequent transits in the light curve. This can cause additional peaks not associated with the rotation period to appear in the ACF or change the shape of the ACF peaks, thus leading to missed detections. To avoid this, we recommend that time series be treated to mask possible transit events before applying the \spinspotter algorithm. 

Finally, in a small number of cases, our algorithm identified a period consistent with the literature value, but rejected it due to being slightly below the selection threshold for one or two parameters. In cases like these, adopting less strict selection criteria would lead to recovery of more of these borderline cases.

%-------------------------------------- Two column figure (place early!)

\subsection{Applications}

In this section we discuss the utility of both the sample of rotating stars we identified in the \tess sample and of the \texttt{SpinSpotter} algorithm more broadly.

The rotating stars we identified
in this paper provide a large collection of fast rotators that will be useful as a basis for future statistical investigations of stellar evolution, young stellar associations, and stellar activity. Our catalog is dominated by rapid rotators due to the upper limit on period sensitivity around 9 days imposed by the short baseline of a \tess sector. As a result, our sample is best suited for science cases involving young stars that have not yet spun down, or stars for which the Skumanich-style age-period relationship is interrupted by some other process that delays the spin down process. In the former case, our sample provides a unique opportunity to develop a holistic view of young, nearby stellar populations across the entirety of the night sky, thanks in large part to the excellent sky coverage of \tess. Gyrochronology models, such as those in \citet{mamajek2008}, indicate that our sample covers much of the period space for stars younger than a billion years. Combined with parallaxes and proper motions from the Gaia mission, this sample may be used to create a fully-realized phase space map of young objects in the night sky, which will help us identify and confirm new members of moving groups and young clusters. This in turn will support studies of gyrochronology in young clusters, which are a cornerstone of constraining gyrchronological relationships since their ages can be determined with a greater degree of certainty than for single stars. Additional members of these groups may be identifiable by their fast rotation periods first and then confirmed with Gaia proper motions and parallaxes.

For older stars that maintain short rotation periods, recent studies have shown that M dwarfs may experience a delayed spin-down process, though the mechanism for it is not understood \citep{Agueros2018}. However, the mechanism driving it is thought to take place quickly, as most of these stars are found either on the rapid rotator or slow rotator track and with few objects in between \citep{newton2016}. \citet{Curtis2020} specifically calls for more M dwarf rotators, particularly ones that are members of young clusters, to be found to use as benchmarks for investigating this phenomenon further. Our sample of \tess rotators contains many more fast rotating M dwarfs than those found by Kepler surveys \citep[e.g.][]{McQuillan2014,Reinhold2020}, and therefore provides a ready made sample for future investigations of this topic.

The second product we present in this paper, the \texttt{SpinSpotter} algorithm itself, also has potential for applicability beyond just the \tess 2-minute cadence sample. \texttt{SpinSpotter} is designed to be flexible and usable on any densely-sampled photometric light curve to detect periodicities of up to one third the baseline of the time series. We posit that many thousands more rotators can be found by applying \texttt{SpinSpotter} to \tess 30-minute cadence targets, though recovery rates may be lower on these targets as the signal to noise on dimmer targets will make detecting periodicities more difficult. \texttt{SpinSpotter} can also be used on Kepler light curves, future photometric missions, and even densely-sampled ground based monitoring of specific stars.

Since \texttt{SpinSpotter} is fast and easily implementable, it is an ideal tool for searching large surveys of stars, such as the whole \tess sample, for targets with specific attributes. By adjusting the thresholds for selection criteria, investigators can quickly identify targets with their desired rotational properties, making it an ideal tool for sifting through large data sets for targets that match a desired science case. Additionally, since \texttt{SpinSpotter} returns a variety of descriptive parameters about the LC, its ACF, and individual peaks in the ACF, investigators may tune their selection criteria to suit a variety of inquiries beyond merely the identification of rotation periods. For example, \citet{basri2021} investigated spot lifetimes from Kepler LCs by using the decay in peak heights in the ACF. This sort of analysis is replicable using \texttt{SpinSpotter} using the $A_k$ parameter, which represents the vertex height of the parabolas fitted to the $k$th peak in the ACF. Similarly, the other diagnostics of features in the ACF can be calculated from the parameters $a_k$, $b_k$, and $c_k$ of the fitted parameter, and, if they can be shown to correspond to physical properties of the star, used to investigate other aspects of the stellar surface. Another interesting application would be to investigate differential rotation, particularly for stars observed during both the \tess primary and extended missions. With a separation of two years between observation periods, most stars will have formed entirely new spot complexes which may occupy different latitudes between the two observation periods. If different rotation periods are identified with high confidence for these two time periods, this may indicate that differential rotation is present and that the star is a good target for a more in depth analysis on the subject.

It is also important to note the limitations of the \texttt{SpinSpotter} algorithm and of this sample of \tess rotators. We caution investigators against using exclusion from this sample as indication of a lack of stellar rotation. Evidence of rotation may be absent due to being overwhelmed by other sources of variability, whether from more complex stellar activity or contamination in the \tess aperture, or the star may simply have no large spot complexes on its photosphere during the relevant observation period. Additionally, stars with rotation periods exceeding one third of the observation baseline have a low chance of recovery, and those that are recovered at such periods have a higher chance of being false positive detections, as described in Section \ref{sec:discussion}. 
%Additionally, any rotation periods that approach half the observation baseline should be treated with caution and further compared against other methods.

\section{Conclusion}\label{sec:conclusion}

We have developed \texttt{SpinSpotter}, an autocorrelation-based algorithm to recover stellar rotation periods from densely sampled photometric data in a robust, automated way. \spinspotter searches for the dominant periodic signal in the ACF of the light curve using the FFT, then fits parabolas to peaks occurring at integer multiples of that period value to characterize the shape of these features in the ACF associated with that period. It uses the average of the vertex locations of the peaks to determine the value of $P_{rot}$, and also requires that the peaks display a characteristic height, width, and goodness of fit to the fitted parabolas. For stars with multiple segments of data, such as \tess sectors or Kepler quarters, we treat both the combined light curve and each segment individually, then require the rotation period be found in a minimum number of segments in order to be accepted. We also perform a check of whether the half-value alias of the rotation period was found instead, and report flags for stars where this may be the case.

Our algorithm is fast, flexible, and needs minimal supervision. It is sensitive to periods up to one third of the baseline of the data, though it may recover periods up to one half the baseline of the data with a higher false positive rate. The characteristic peak parameters it reports allow investigators to tune their selection criteria for finding periods to best match their science case, and also opens the door to perform other types of photometric activity analysis beyond just identifying the value of the rotation period, such as searching for the presence of differential rotation or the decay times of spot complexes.

We applied our algorithm to $\sim$130,000 main sequence stars observed by \tess at 2-minute cadence and found rotation periods for 13,504 stars ranging from 0.4 to 14 days. We compared our results to surveys of known rotators from the Kepler sample, the \tess TOI list, and members of young clusters and found good agreement with the literature. Because we are limited to fast rotators by the short 27-day baseline of a \tess sector, our sample is most useful for studying young stars, particularly those less than $\sim$1 billion years old, or those that undergo a delayed spin down process. The latter case is particularly interesting, since we identify a large population of fast-rotating M dwarfs not present in previous Kepler rotation studies. These stars may make an excellent basis for future studies into the mechanisms that govern the evolution of stellar spin down in late type stars. We also see utility for this data set in searching for previously unidentified members of nearby young stellar associations. Rapid rotation, combined with proper motion and parallax data from the Gaia mission, may provide strong evidence for membership of individual stars in these groups.

Though not without its challenges, the \tess mission gives us an unprecedented opportunity to investigate patterns of stellar rotation across nearly the entirety of the night sky. Through the \texttt{SpinSpotter} algorithm and our sample of \tess rotators, we aim to provide tools for future investigations into stellar rotation and evolution.

\section{Acknowledgements}\label{sec:Acknowledgements}

This paper includes data collected by the TESS mission. Funding for the TESS mission is provided by the NASA's Science Mission Directorate.

\bibliographystyle{aasjournal} 
\bibliography{bib}

\end{document}